\documentclass[aps,prl,preprint,superscriptaddress]{revtex4-1}
\usepackage[usenames]{color}
\usepackage{epsf}
\usepackage{epsfig}
\usepackage{graphicx}
\usepackage{latexsym}
\usepackage{bm}

\begin{document}
~\\
published in: Phys. Rev. Lett. 104, 153401 (2010)
~\\
\title{Single photon double ionization of the helium dimer}

\author{
T. Havermeier$^{1}$, T. Jahnke$^{1}$, K. Kreidi$^{1}$, R.
Wallauer$^{1}$, S. Voss$^{1}$, M. Sch\"offler$^{1}$, S.
Sch\"ossler$^{1}$, L. Foucar$^{1}$, N. Neumann$^{1}$, J.
Titze$^{1}$, H. Sann$^{1}$, M. K\"uhnel$^{1}$, J.
Voigtsberger$^{1}$, A. Malakzadeh$^{1}$, N. Sisourat$^{2}$, W.
Sch\"ollkopf$^{3}$, H. Schmidt-B\"ocking$^{1}$, R. E.
Grisenti$^{1,4}$ and R. D\"orner$^{}$}


\address{
 Institut f\"ur Kernphysik, J.~W.~Goethe Universit\"at, Max-von-Laue-Str.1, 60438 Frankfurt, Germany \\
$^2$ Institut f\"ur physikalische Chemie, Universit\"at Heidelberg, Im Neuenheimer Feld 229, 69120 Heidelberg, Germany\\
$^3$ Fritz-Haber-Institut der Max-Planck-Gesellschaft, Faradayweg 4-6, 14195 Berlin, Germany\\
$^4$ GSI Helmholtzzentrum f\"ur Schwerionenforschung GmbH Planckstr.
1, 64291 Darmstadt, Germany}

\begin{abstract}
We show that a single photon can ionize the two helium atoms of the
helium dimer in a distance up to 10 {\AA}. The energy sharing among
the electrons, the angular distributions of the ions and electrons
as well as comparison with electron impact data for helium atoms
suggest a knock-off type double ionization process. The Coulomb
explosion imaging of He$_2$ provides a direct view of the nuclear
wave function of this by far most extended and most diffuse of all
naturally existing molecules.
\end{abstract}
\maketitle

The helium dimer ($^{4}$He$_{2}$) is an outstanding example of a
fragile molecule whose existence was disputed for a long time
because of the very weak interaction potential\cite{Tang1995PRL}
(see black curve in Fig. 1). Unequivocal experimental evidence for
$^{4}$He$_{2}$ was first provided in 1994 in diffraction
experiments\cite{Schoellkopf1994Science} by a nano-structured
transmission grating. Subsequently the average dimer bond length and
dimer binding energy could be determined to be 52 {\AA} and
10$^{-7}$ eV (0.9$\cdot$10$^{-3}$ cm$^{-1}$ or 1.3
mK)\cite{Grisenti2000PRL}. This very large bond length, a factor 100
larger than the hydrogen bond length, goes along with a prediction
of very widely delocalized wave function, unseen in any other
molecule \cite{Luo96} (see blue function in Fig. 1). It is because
of these exotic properties that "as the hydrogen molecule in the
past, the helium dimer today became a test case for the development
of new computational methods and tools"\cite{Komasa2001JCP} in
quantum chemistry. Despite this fundamental nature of the diffuse
helium dimer wave function, it has escaped direct experimental
observation until now, as the diffraction grating experiment measure
the mean value and not the shape of the wave function itself. Our
experiment provides a direct view of this diffuse object.

The large distance between the two helium atoms and the minuscule
binding energy make the Helium dimer an unique model system to
explore electron correlations over large distances. The most
sensitive tool for such studies is multiple photoionization. Since
photoabsorption is described by a single electron operator the
photon energy and angular momentum is best thought of as being
initially given to one electron of the atom only. In the absence of
electron correlation the ejection of a single electron would be the
only possible outcome of the photoabsorption process. Due to the
ubiquity of electron correlation, however, the ejection of electron
pairs by a single photon is a wide spread phenomenon seen in
atoms\cite{briggs2000mop},
molecules\cite{weber2004nature,vanroose2005science} and
solids\cite{Kouzakov2003PRL}. This two electron process poses at
least two central questions: what is the correlation mechanism by
which the photon energy is distributed among the two electrons and
over which distance are such correlations active? In the present
work we report the surprising observation that a single photon can
lead to nonsequential ejection of two electrons from two atoms which
are separated by many atomic radii and where the overlap of the
electronic wave functions is negligible. By measuring the
internuclear distance for each ionization event together with the
emitted electron pair, we show that as much as 39.3 eV of energy can
be transferred up to a distance of 10 {\AA}. The energy is
transferred by a type of ''interatomic billiard''. Our measured
angular distributions suggest that primarily one electron absorbs
the energy and is ejected from one of the atoms of the dimer. In
some cases it is emitted towards the neighboring atom where it
transfers part of its energy to a second electron which is knocked
off.

For our experimental investigation we employed COLTRIMS momentum
imaging\cite{doerner00pr,Ullrich2003RPP,Jahnke04JESRP}. A supersonic
jet is crossed with a linearly polarized photon beam inside a
spectrometer at beamline UE112PGM2 of the BESSY synchrotron.. By
means of homogenous electric (E = 12 V/cm) and magnetic fields (B =
10 G) all charged particles created in the reaction are mapped
towards two position and time sensitive
detectors\cite{Jagutzki2002NIMP}. By measuring the positions of
impact and the times-of-flight of all particles in coincidence the
vector momentum of each particle is obtained during
offline-analysis.  In order to create Helium dimers we have expanded
Helium gas through a 5 $\mu$m nozzle cooled to a temperature of 18
K. A driving pressure of 1.8 bar and a pressure of
1.2$\cdot$10$^{-4}$ mbar on the low pressure side of the nozzle
yielded a dimer fraction of 1-2 \% in the gas beam.

If both He atoms of a dimer are ionized the dimer will undergo
Coulomb explosion. In that case both He$^{+}$ ions are emitted
back-to-back with a momentum of same magnitude. This is a clear
signature for the breakup of a dimer, thus reactions of trimers and
lager clusters can be excluded from the dataset. In the absence of
electron correlation no He$^{+}$-He$^{+}$ coincidences would be
expected. Therefore already the observation of He$^{+}$-He$^{+}$
coincidences proves experimentally the electron correlation that
exists between electrons from the two distant atomic centers of the
dimer. It is able to transfer at least 24.59 eV of energy (which
equals the ionization energy of Helium) across the bond length of
the dimer.

\begin{figure}[ht]
  \begin{center}
  \includegraphics[width=14.0 cm]{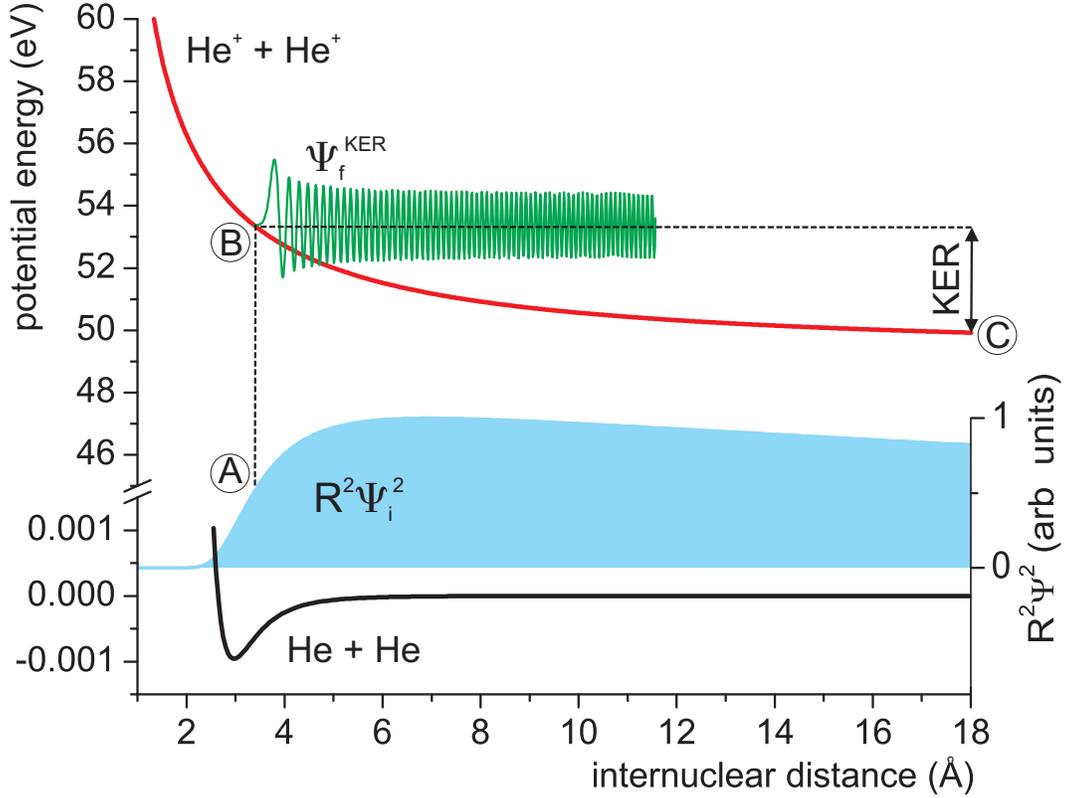}
  \caption{The distribution of internuclear distances R (R$^{2}\Psi_{i}^{2}$(R)) of
  the helium dimer (blue) is spread over a range up to several
  hundred Angstrom (out of scale). Via photoionization a transition from the ground state (black) to
  the repulsive potential of two He-ions (red) is possible (path A to B).
  In the classical reflection approximation, dissociation (path B to C) leads
  to a certain kinetic energy release (KER) of the two ions.
  In a quantum mechanical description the one-to-one relation of R and KER
  along path A-B-C is replaced by the overlap of $\Psi_i$(R) with the final state
  wave function $\Psi_{f}^{KER}$(R) (green) for the respective KER.}
  \end{center}
\end{figure}

\begin{figure}[ht]
  \begin{center}
  \includegraphics[width=8.0 cm]{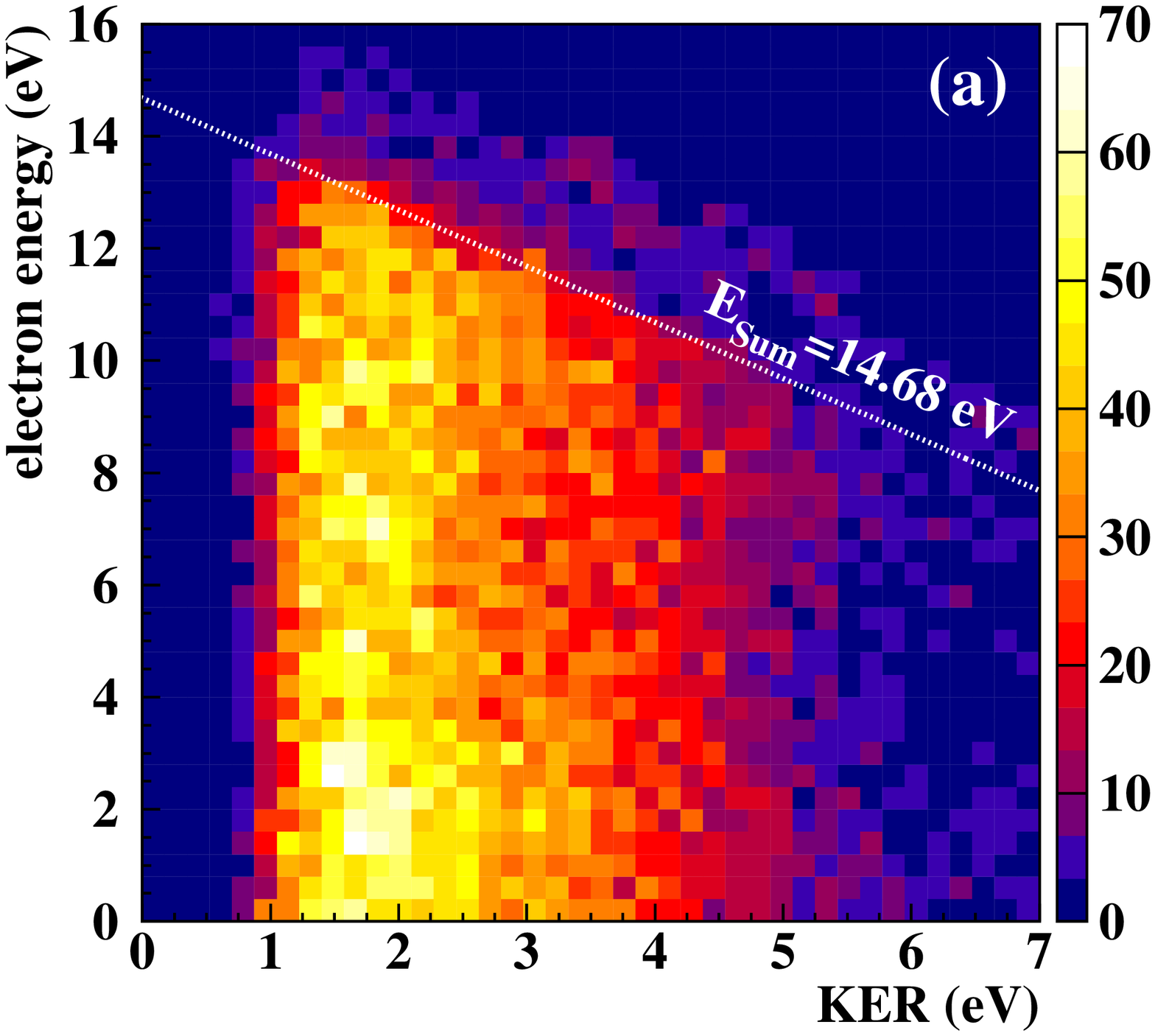}
  \includegraphics[width=8.0 cm]{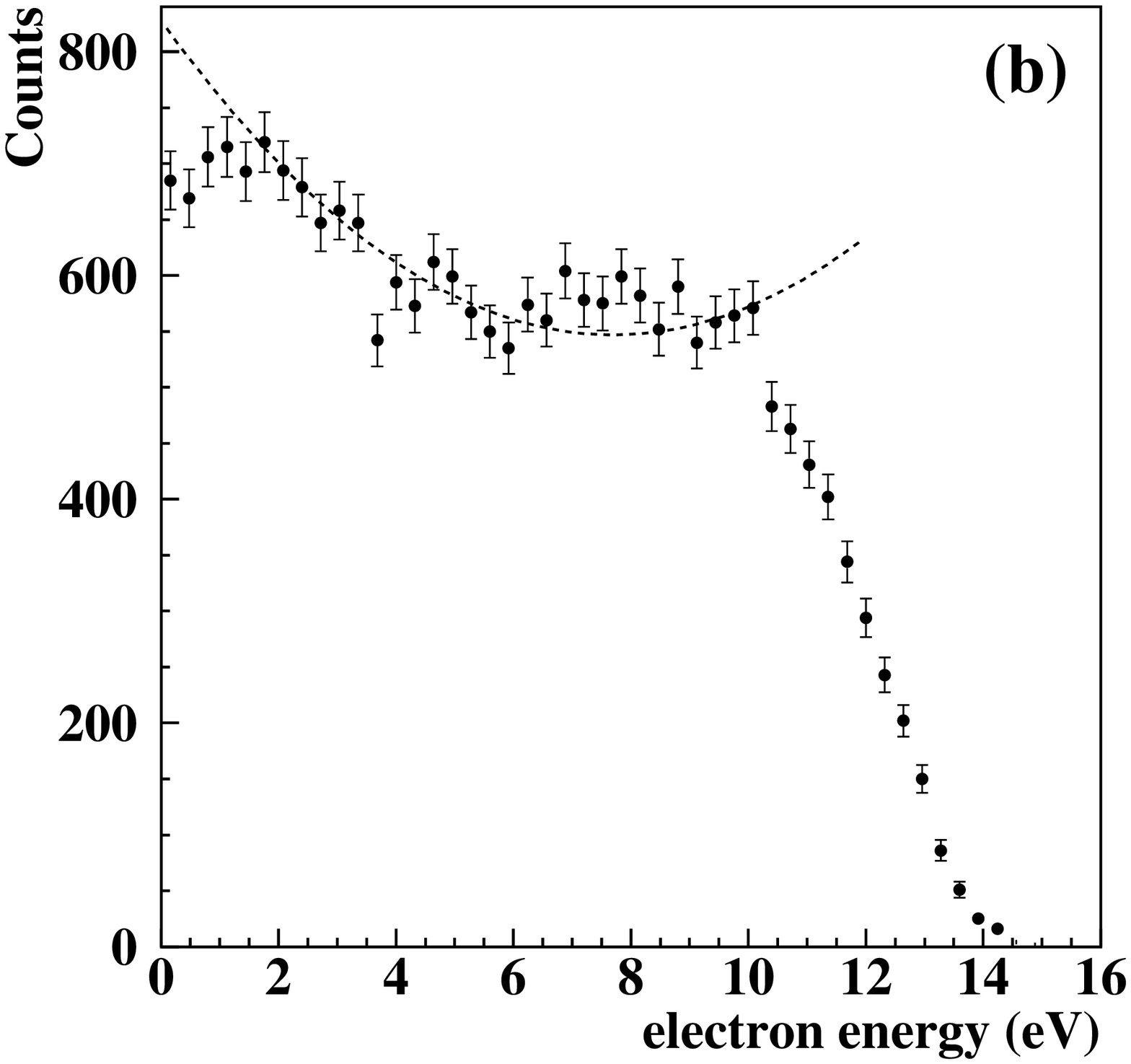}
  \caption{Experimental results for double ionization of He$_{2}$ by one
  63.86 eV linearly polarized photon. (a) kinetic energy release (KER) of
  both fragment ions versus energy of one of the two electrons.
  The dashed line at a sum energy of 14.68 eV is given by the total available energy.
  Data points below 1.5 eV KER are affected by decreased detection efficiency.
  (b) Electron energy spectrum (projection of (a) on vertical axis).
  Dashed line: CCC theory, electron impact ionization (e,2e) of He at 40eV initial energy \cite{bray2003}.}
  \end{center}
\end{figure}

\begin{figure}[ht]
  \begin{center}
  \includegraphics[width=7 cm]{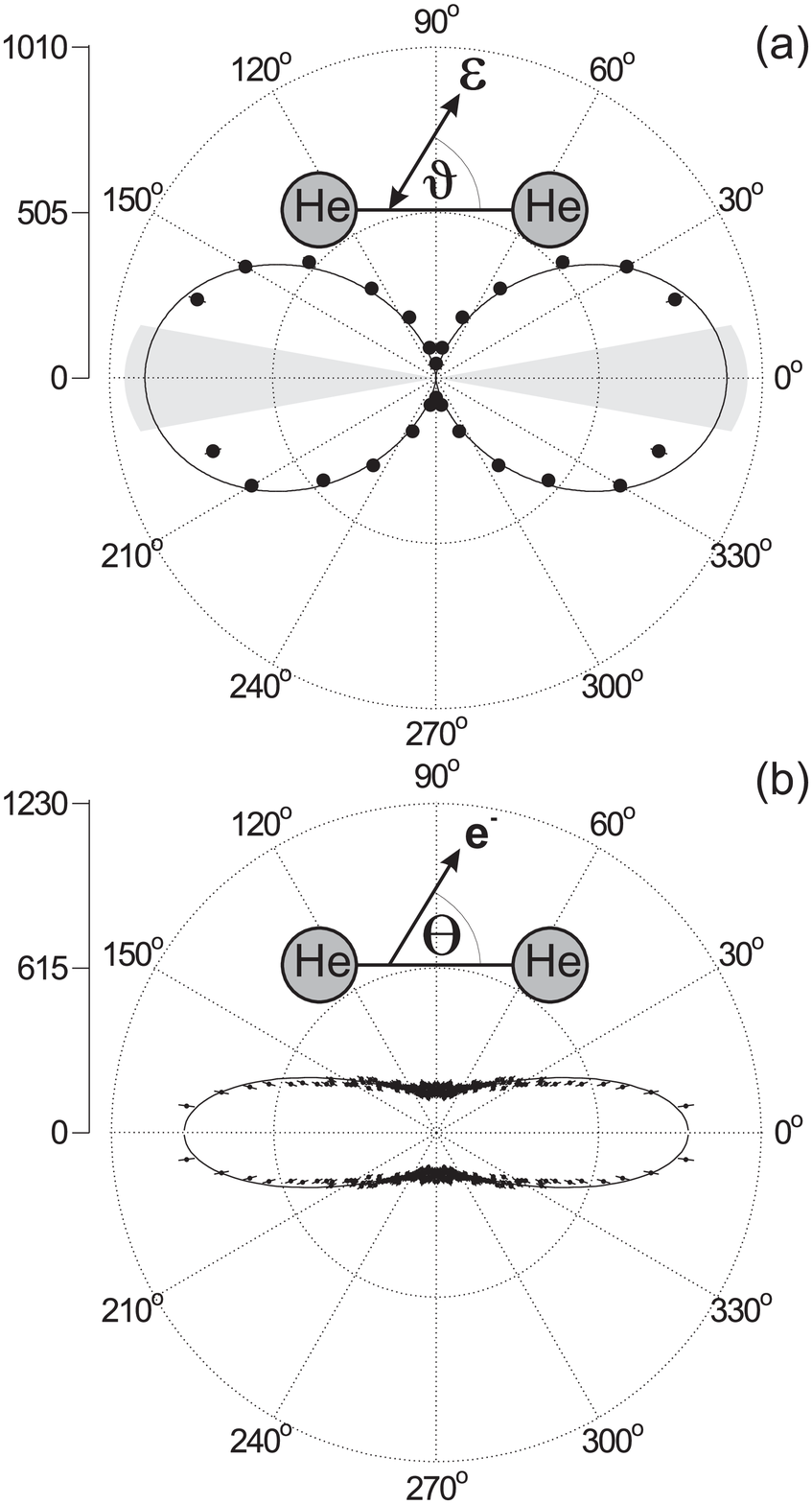}
  \caption{Upper panel: angle of the dimer axis with
  respect to the photon polarization $\varepsilon$ (horizontal), the black line shows
  a dipole distribution fitted to the data ($\cos^2(\vartheta)$).
  Some data points are missing due to a decreased detection efficiency
  for certain regions of angles and KERs (gray area).
  Lower panel: angle between the dimer axis and outgoing electrons
  integrated over all electron energies.
  Continuous line: CCC calculation for electron impact ionization (e,2e) of He at 40eV initial energy.
  The calculation is integrated over the secondary electron energies at 1, 4, 7.7, 12.7 and 15 eV
  from \cite{bray2003}. The theory has been mirrored
  to account for the indistinguishability of the two centers in the helium dimer.}
  \end{center}
\end{figure}

To unravel the energy transfer mechanisms we now examine how the
photon energy is distributed among the two electrons and two ions.
Fig. 2a shows the sum of the energy of both of the He$^{1+}$ ions
that were emitted back-to-back (kinetic energy release KER) versus
the energy of one of the electrons measured in coincidence. The
ionization potential of the He atom is E$_{iHe}$=24.59 eV. At a
photon energy of E$_{photon}$=63.86 eV a total of E$_{photon}$-
2$\cdot$E$_{iHe}$ = 14.68 eV is left for the four particles to
share. The diagonal line in Fig. 2a shows this upper bound.

Fig. 2b shows the energy spectrum of the emitted electrons. Between
0 and 10 eV, the distribution is slightly declining, followed by a
steep decay. The first region is similar to the energy sharing of
two electron emitted in direct single photon double ionization of a
single Helium atom\cite{Wehlitz1991PRL} or a covalently bound
molecule\cite{doerner98prl}, suggesting similar correlation
mechanisms to be at work. For the long studied atomic case knock-off
and shake-off are the two well established mechanisms mediating
double electron
ejection\cite{Knapp2002PRL,Hino2001PRA,Kheifets2001JPB,Schneider2002prl}.
In the knock-off process the energy is transferred by a binary type
collision between the two electrons. For shake-off the rapid change
of the correlated wave function caused by the photo ejection of one
electron leads to shake off of the second electron upon the
relaxation of this wave function to the new eigenstates of the
altered potential. In the case of a knock-off process, the second
ionization step can be considered as the impact of a 39.3 eV
electron on a single helium atom (compare
\cite{Samson90prl,Schneider2002prl}). This (e,2e) type process has
been examined in multiple experiments (i.e.
\cite{Ehr1971,Röder1997,SchowPRL2005}) and is well understood by
theory. The dashed line in Fig. 2 shows a convergent close-coupling
(CCC) calculation of the single differential cross section, i.e. the
electron energy distribution, for an (e,2e) collision of a free 40
eV electron with a neutral He atom. It agrees surprisingly well with
our measured distribution for photo double ionization of He$_2$. The
edge above 10 eV is due to restricted sum energy. In contrast to an
(e,2e) experiment, the amount of 14.68 eV is not only shared by two
electrons, but also by the ions (KER).

The angular distribution of the dimer axis with respect to the
photon polarization (Fig. 3a) allows to distinguish between
shake-off and knock-off. Shake-off is polarization independent
resulting in an isotropic distribution of the ionic fragments in the
laboratory frame. For the knock-off however a simple billiard type
scenario suggests that the photoelectron which is launched at one
side can hit the second atom only if the dimer axis is oriented
along the direction of the primary photoelectron momentum. If that
is the case, the angular distribution of the dimer axis should
reflect the angular distribution of the photoelectron of the He
monomer. For the case of ionization of an 1s state as in helium this
is a dipolar distribution given by $\cos^2(\vartheta)$,where
$\vartheta$ is the angle between the photoelectron and the
polarization axis. The measured experimental angular distribution of
the dimer axis (Fig. 3a) resembles strikingly that distribution and
thus gives a clear evidence for the knock-off process ruling out the
shake off mechanism.

As we have argued above, the primary photoelectron has to be emitted
approximately along the dimer axis to facilitate the knock-off at
the neighbor. Thus in the present case the dimer axis is the
equivalent to the electron beam axis in the (e,2e) experiment at an
atom. We therefore compare angular distributions of the electron
with the dimer axis in the present experiment with the electron
angular distribution with respect to the beam axis in the (e,2e)
case (Fig. 3b). As it is not possible to distinguish from which
atomic center the photoelectron is emitted, the intensities
I($\theta$) and I(180-$\theta$) add. Hence the data of the CCC
calculation \cite{bray2003} for the (e,2e) collision with He has
been mirrored accordantly. The theory (line in Fig. 3b) is in
excellent agreement with the measured distribution.

Our identification of the knock-off mechanism as being responsible
for the observed single photon double ionization is hence supported
by three independent experimental observables: the energy sharing
between the electrons, the angular distribution of the molecular
axis and the angular distribution of the electrons with respect to
the molecular axis.

\begin{figure}[ht]
  \begin{center}
  \includegraphics[width=10 cm]{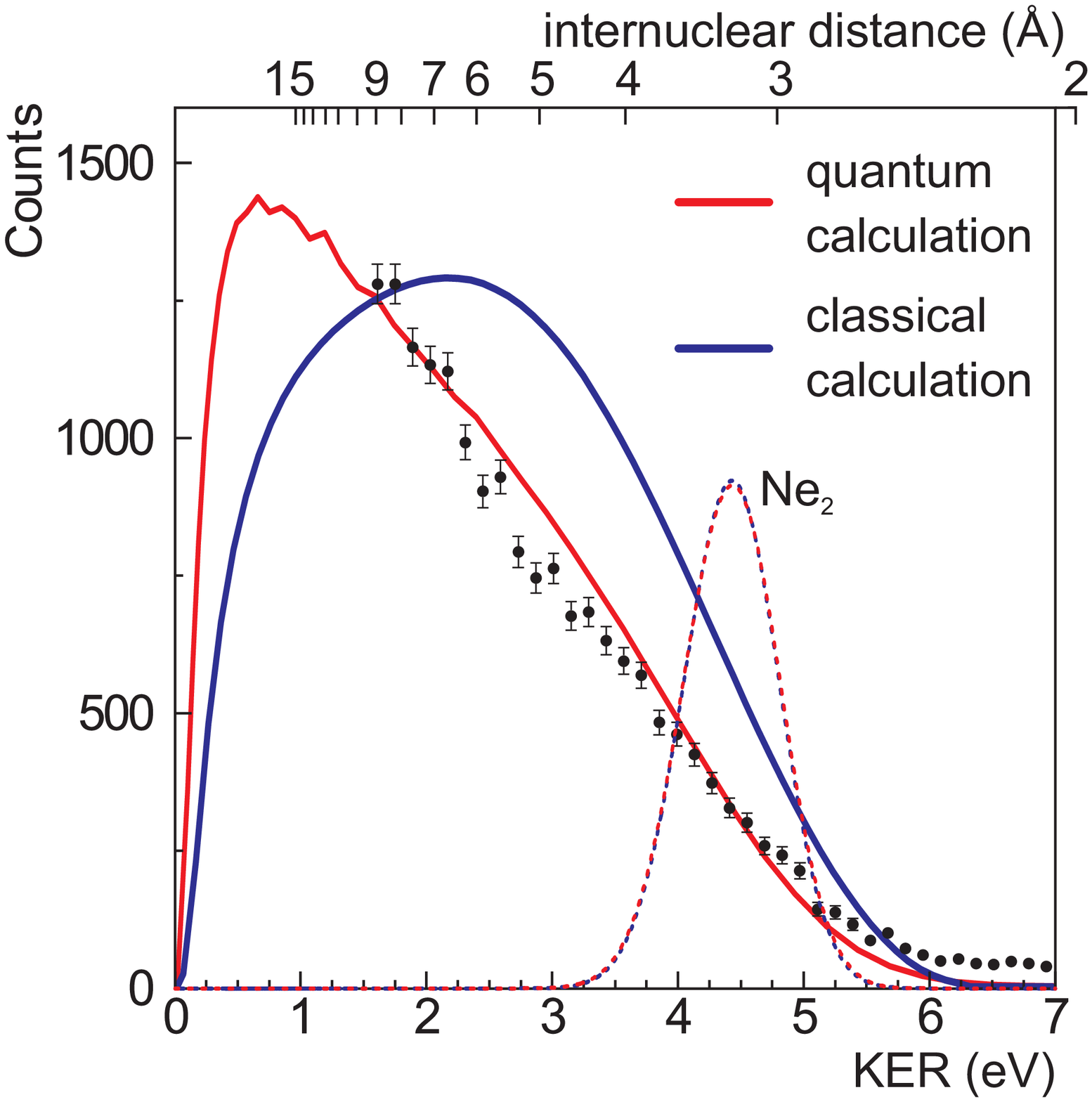}
  \caption{KER distribution. Blue line: classical calculation (reflection approximation,
  path A-B-C in Fig. 1) red line: quantum calculation. Data points below 1.5 eV are missing
  due to a decreased detection efficiency in this area.
  The doted lines show the two calculations for the neon dimer.}
  \end{center}
\end{figure}

We now use this knowledge of the correlation mechanism and the
measured coulomb explosion energy to sample the He$_{2}$ ground
state wave function. This can be done in full quantum version or in
a classical mechanics approximation. Within the realm of classical
mechanics there is a one to one relation of the internuclear
distance R of the dimer at the instant of double ionization (labels
on top of Fig. 4) to the KER (labels on bottom of Fig. 4).  As
illustrated by the dashed line in Fig. 1 photon absorption promotes
the system from the ground state to the He$^+$+He$^+$ potential
(path A to B). Dissociation on this repulsive curve (path B to C)
leads to KER=C/R with the proportionality constant C=52eV/{\AA}. The
classical physics reflection
approximation\cite{weber2004nature,Gislason1973JCP} is successfully
used e.g. in Coulomb explosion imaging in ion
beams\cite{Vager1989science} and strong laser
fields\cite{Ergler2005PRL}.

Quantum mechanically the classical relationship KER = 1/R has to be
replaced by the overlap of the bound initial state wave function
$\Psi_i$(R) of the dimer with the continuum wave function
$\Psi{_f}^{KER}$(R) for the respective KER (see Fig. 1). The
probability distribution P(KER) is then given by:
\begin{equation}
P(KER)=\left|\int dR\Psi_i(R)\frac{1}{R}\Psi_f^{KER}(R)\right|^2.
\end{equation}
Since the second (target) atom can only be ionized if the
photoelectron is emitted within a certain solid angle, the knock-off
correlation mechanism underlies a classical 1/R$^{2}$ dependency. In
the quantum treatment accordingly a geometrical factor 1/R has to be
inserted to the integral of eq. 1. Classical and quantum sampling of
the ground state wave function are shown by the blue and red line in
Fig. 4. While the classical approximation introduces only  a small
error when the reflecting potential is very steep and the initial
state is confined to a narrow region of R as in covalently bound
molecules\cite{weber2004nature} the extreme dimensions of the Helium
dimer require the full quantum treatment. The good agreement of  the
present coulomb explosion data with the full quantum calculation
confirms the unique delocalized character of the He$_2$ wave
function as predicted by theory. Our experiment samples this wave
function in the region of approximately 2.5-10 Angstrom. We note
that the typical confined gaussian wave function of all standard
ground state molecules and also of all other van-der-Waals Systems
would lead to a near gaussian peak in the KER. For comparison singly
photon double ionization of neon dimers leads to a near Gaussian
peak as narrow as $\Delta$E/E= 0.2 peaked at 4.4 eV. The differences
between classical and quantum calculation is negligible in Ne$_2$
(see dotted lines in Fig 4).

In conclusion we have observed the electron correlation mediated
direct ejection of two electrons from two distant sites upon
absorption of a single photon. The underlying mechanism of a
knock-off type process which has its analogy in double ionization of
atoms. For the extreme conditions of the helium dimer the knock off
can be split into a photoionization at one center (the photoelectron
gun) followed by an e,2e collision at the neighbor. This e,2e
collision occurs in the static field of the neighboring coulomb
charge, which can be controlled by the internuclear distance. This
distance also controls the phase of the photoelectron as it hits the
neutral. With very strong external field and phase, two new control
parameters are introduced in e,2e. The influence of these parameters
escaped observation and also theoretical attention so far. It is
related to the rescattering process in a strong laserfield, where
the intermediate electron is additionally driven by the field.

\acknowledgments  We want to thank the staff of BESSY II for
experimental support. This work was funded by the Koselleck project
of the Deutsche Forschungsgemeinschaft. R. E. G. and M. K.
acknowledge support by the Helmholtz society, grant VH-NG-331. We
acknowledge helpful discussions with Igor Bray.

\end{document}